\documentclass{article}

\usepackage{PRIMEarxiv}

\usepackage[utf8]{inputenc} 
\usepackage[T1]{fontenc}    
\usepackage{hyperref}       
\usepackage{url}            
\usepackage{booktabs}       
\usepackage{amsfonts}       
\usepackage{nicefrac}       
\usepackage{microtype}      
\usepackage{lipsum}
\usepackage{accessibility} 
\usepackage{fancyhdr}       
\usepackage{graphicx}       
\usepackage{amsmath}
\usepackage{hyperref}
\usepackage{colortbl} 
\usepackage[dvipsnames,table]{xcolor} 
\usepackage{multirow} 
\usepackage{float}
\graphicspath{{media/}}     

\title{SkinningGS: Editable Dynamic Human Scene Reconstruction Using Gaussian Splatting Based on a Skinning Model
\thanks{\textit{\underline{Citation}}: 
\textbf{Authors. Title. Pages.... DOI:000000/11111.}} 
}

\author{
  Da Li, Donggang Jia, Markus Hadwiger, Ivan Viola \\
  \\
  King Abdullah University of Science and Technology \\
  \\
  \texttt{\{da.li, donggang.jia,markus.hadwiger,ivan.viola\}@kaust.edu.sa} \\
}

\begin{document}
\maketitle

\begin{abstract}
Reconstructing an interactive human avatar and the background from a monocular video of a dynamic human scene is highly challenging. In this work we adopt a strategy of point cloud decoupling and joint optimization to achieve the decoupled reconstruction of backgrounds and human bodies while preserving the interactivity of human motion. We introduce a position texture to subdivide the Skinned Multi-Person Linear (SMPL) body model's surface and grow the human point cloud. To capture fine details of human dynamics and deformations, we incorporate a convolutional neural network structure to predict human body point cloud features based on texture. This strategy makes our approach free of hyperparameter tuning for densification and efficiently represents human points with half the point cloud of HUGS. This approach ensures high-quality human reconstruction and reduces GPU resource consumption during training. As a result, our method surpasses the previous state-of-the-art HUGS in reconstruction metrics while maintaining the ability to generalize to novel poses and views. Furthermore, our technique achieves real-time rendering at over 100 FPS, $\sim$6$\times$ the HUGS speed using only Linear Blend Skinning (LBS) weights for human transformation. Additionally, this work demonstrates that this framework can be extended to animal scene reconstruction when an accurately-posed model of an animal is available.
\end{abstract}


\begin{figure}
\centering
  \includegraphics[width=0.9\textwidth]{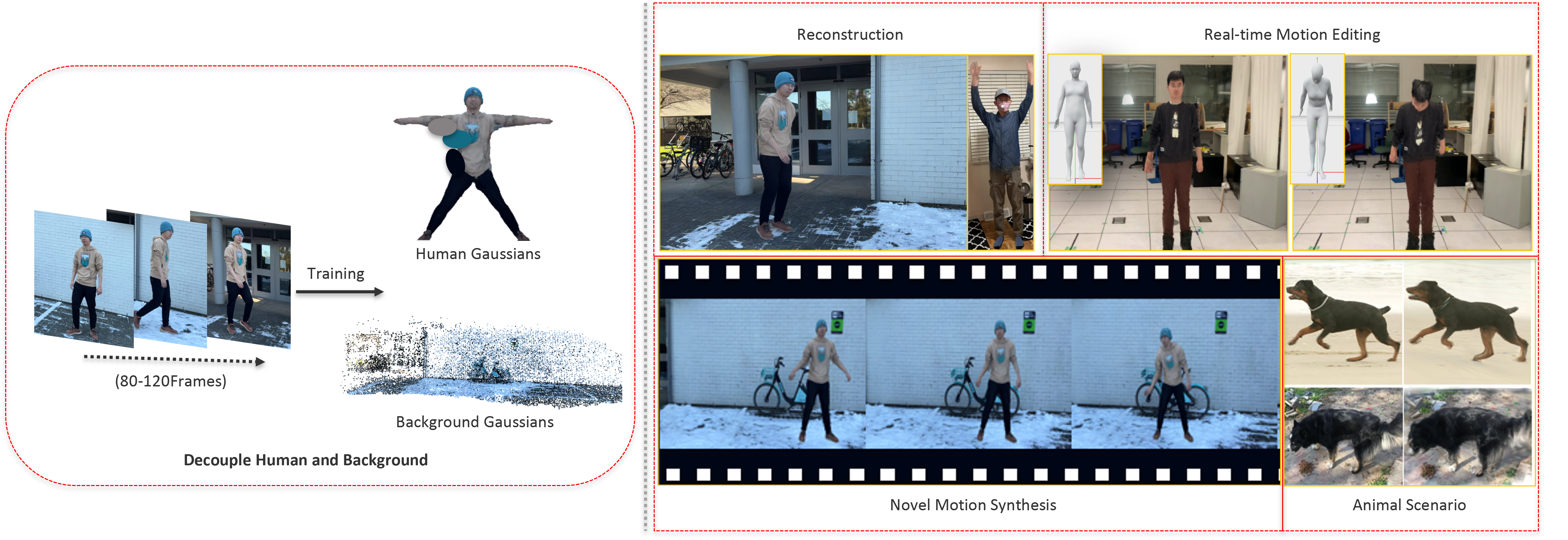}
  \caption{Our method can decouple a human from the background given a monocular video. The reconstruction preserves interactive capacity, which can be used to develop novel motion synthesis and other applications. Furthermore, our method can be extended for animal scene reconstruction.}
  \label{fig:teaser}
\end{figure}

\section{Introduction}
Scene reconstruction technology plays a pivotal role in applications such as virtual reality (VR), augmented reality (AR), film production, 3D asset modeling, and virtual try-on, enabling the digitization of real-world scenes and providing immersive visual experiences. However, reconstructing dynamic scenes, particularly those involving humans, while preserving interactivity remains a significant challenge.

This work addresses the disentanglement and reconstruction of humans and backgrounds from monocular videos while maintaining human motion interactivity. Recent advancements in Neural Radiance Fields (NeRF)~\cite{mildenhall2021nerf} have revolutionized static scene reconstruction by representing scenes as implicit functions, enabling high-quality novel-view synthesis~\cite{xu2022point,barron2021mip}. Extensions like HyperNeRF~\cite{park2021hypernerf} adapt NeRF for dynamic scenes through temporal deformation fields~\cite{li2021neural,li2022neural,park2021nerfies}. Despite these innovations, NeRF-based methods face challenges such as low training efficiency and limited rendering speed for dynamic scenes.

In contrast, explicit representations like 3D Gaussian point clouds~\cite{kerbl20233d} have emerged as efficient alternatives. These methods achieve high rendering efficiency by leveraging sparse Gaussian point clouds without compromising visual quality. Recent developments in Gaussian-based methods, particularly Gaussian splatting, have advanced dynamic scene reconstruction by incorporating temporal deformation fields~\cite{yang2024deformable,lin2024gaussian,guo2024motion,yang2023real} or physically-based priors~\cite{luiten2023dynamic}. However, both NeRF-based and Gaussian-based approaches generally sacrifice interactivity and editability.

ST-NeRF~\cite{Zhang_2021} addresses these limitations by introducing a multi-layer network representation to enhance interactivity and editability. However, it relies on multi-view inputs and lacks a human prior, limiting its ability to generate novel motions. Neuman~\cite{jiang2022neuman} further enables editable dynamic human reconstruction from monocular videos, but inherits the slow rendering and training speeds of NeRFs, restricting real-time applications. More recently, HUGS~\cite{kocabas2024hugs} introduced a Gaussian-based representation for real-time rendering, partially overcoming the limitations of Neuman. However, HUGS suffers from several drawbacks: 1) its three-plane network fails to capture fine surface details; 2) its deformation field and densification strategies cause distortions at human-background overlaps; and 3) its point cloud growing depends heavily on scene-specific hyperparameter tuning.

To enhance human surface detail capture, we propose a novel approach that introduces position texture mapping to enrich the SMPL~\cite{loper2015smpl} model’s base vertices, enabling more detailed human point cloud representation. Additionally, we incorporate the PoP (Power of Points) network architecture~\cite{hu2024gaussianavatar,ma2021power, SCALE}, which uses texture features to predict 3D Gaussian properties in canonical space. This allows our method to represent humans with fewer points, achieving faster rendering speeds. Our main contributions are:
\begin{itemize}
\item We introduce a point cloud growing method based on a position texture to improve the expressiveness and detail capture capability of the human representation.
\item We demonstrate that a point cloud feature predictor based on texture mapping enables better reconstruction results in joint optimization with the background.
\item We show that our method surpasses HUGS and Neuman on two benchmarks regarding reconstruction quality while achieving $\sim$6$\times$ the speed of HUGS, allowing real-time rendering even on low-powered GPUs.
\item We demonstrate how our method can be extended to animal scenes and real-time motion editing applications.
\end{itemize}

\section{Related Research}
We introduce general scene reconstruction techniques, followed by NeRF-based and Gaussian Splatting-based methods, and conclude with related works on reconstructing dynamic human scenes.
\subsection{Scene Representation and Rendering}
Scene reconstruction and rendering have evolved from early 4D light representations~\cite{gortler1996lumigraph,levoy2023light}, which enabled efficient multi-view rendering but demanded dense data sampling, to geometry-based methods like Structure-from-Motion (SfM)~\cite{snavely2006photo}. SfM employs multi-view geometry and sparse point cloud generation, with later improvements~\cite{schonberger2016structure,schonberger2016pixelwise} enhancing robustness and scalability. Despite these advances, point cloud reconstruction still struggles with fitting issues~\cite{kerbl20233d}, limiting its applicability in complex scenes.

NeRF~\cite{mildenhall2021nerf} revolutionized scene reconstruction by using implicit 5D radiance fields to generate high-quality views from limited inputs. Early extensions like Mip-NeRF~\cite{barron2021mip} and Point-NeRF~\cite{xu2022point} improved sampling efficiency and detail capture. For dynamic scenes, methods incorporating temporal deformation fields~\cite{park2021nerfies,li2021neural} or high-dimensional extensions like HyperNeRF~\cite{park2021hypernerf} enabled modeling topological changes. More advanced methods~\cite{li2022neural,li2023dynibar} integrated spatiotemporal sampling and motion-aware features for better dynamic reconstruction. However, these approaches face limitations in real-time rendering and interactivity.

Gaussian Splatting (3DGS)~\cite{kerbl20233d} offers an explicit alternative to NeRF by using sparse Gaussian point clouds, achieving efficient rendering without compromising quality. Recent advancements~\cite{yang2024deformable,lin2024gaussian,guo2024motion,yang2023real} have adapted 3DGS for dynamic scenes using temporal deformation fields or optical flow features, while physically consistent constraints~\cite{luiten2023dynamic} enhance motion capture. However, issues like dependence on multi-camera inputs and limited editability persist. Despite accelerating dynamic scene reconstruction, these methods struggle with editable and interactive dynamic human motion, limiting their applicability for reconstructing dynamic human bodies.
\subsection{Human Avatar Reconstruction}
Parameterized models like SMPL~\cite{loper2015smpl} and its extensions, such as SMPL-X~\cite{pavlakos2019expressive}, laid the foundation for human avatar reconstruction by combining linear blend skinning with semantic refinement~\cite{alldieck2018detailed}. More recent methods like Vid2Actor~\cite{weng2020vid2actor} and HumanNeRF~\cite{weng2022humannerf} introduced skeleton-driven deformation and volumetric rendering for dynamic avatars. Gaussian-based techniques~\cite{qian20243dgs,hu2024gaussianavatar,li2024animatable} have improved efficiency and motion detail, but remain limited by challenges in integrating humans with backgrounds.

\subsection{Scene-Human Joint Optimization}
Joint optimization with decoupled representations for background and humans has proven effective for high-quality reconstruction and editability. ST-NeRF~\cite{Zhang_2021} uses layered neural representations and 4D label maps for spatial and temporal decoupling, enabling rich editing capabilities like scaling and timeline adjustments. However, its reliance on 16 RGB cameras limits scalability. Neuman~\cite{jiang2022neuman} employs dual NeRF models with SMPL priors and an error-correction network for dynamic effects, but suffers from slow optimization and rendering speeds. HUGS~\cite{kocabas2024hugs} introduces animated 3D Gaussian point clouds driven by LBS for editable dynamic modeling. While effective, HUGS faces challenges with deformation artifacts, texture blurring, and reliance on hyperparameter tuning, limiting its generalizability.

Our method maintains a decoupled representation of background and human Gaussian point clouds through a growing strategy based on a position texture. By employing a texture-driven PoP network, we efficiently predict 3D Gaussian point cloud features, enabling more detailed human representation with fewer points and accelerated rendering speeds.

Table~\ref{tab:comparison} summarizes a comparison of these methods, highlighting their key differences in functionality. While ST-NeRF effectively decouples humans and scenes, it fails to support novel human motion. Neuman can synthesize novel motion but suffers from slow rendering speeds. Our method outperforms HUGS in reconstruction quality. Moreover, our real-time rendering speed is $\sim$6$\times$ of HUGS, enabling real-time performance even on less powerful GPUs.

\begin{table}[h!]
    \centering
    \small
    \begin{tabular}{@{}lcccp{4cm}@{}}
        \toprule
        \textbf{Methods} & \textbf{Real-time} & \textbf{Novel motion} & \textbf{Human-scene decoupling} \\
        \midrule
        ST-NeRF                 & \(\times\)        & \(\times\)            & \(\checkmark\) \\
        Neuman                  & \(\times\)        & \(\checkmark\)        & \(\checkmark\) \\
        HUGS                    & nearly    & \(\checkmark\)        & \(\checkmark\) \\
        Ours                    & \(\checkmark\)    & \(\checkmark\)        & \(\checkmark\) \\
        \bottomrule
    \end{tabular}
    \caption{Comparison of the functionality of different methods.}
    \label{tab:comparison}
\end{table}

\section{Method}

\subsection{Preliminaries}

\subsubsection{Gaussian Splatting}
3D Gaussian Splatting~\cite{kerbl20233d} is a framework for 3D reconstruction using point clouds with Gaussian kernels. Each point is represented by a Gaussian centered at \( \mathbf{p} = (x, y, z) \) with a covariance matrix \( \mathbf{V} \). The Gaussian kernel is defined as:
\begin{equation}
G_{\mathbf{V}}(\mathbf{x} - \mathbf{p}) = \frac{1}{(2\pi)^{3/2} |\mathbf{V}|^{1/2}} \exp\left(-\frac{1}{2} (\mathbf{x} - \mathbf{p})^\top \mathbf{V}^{-1} (\mathbf{x} - \mathbf{p})\right).
\end{equation}

Each point has attributes such as transparency (\( \alpha \)) and RGB color. The rendering process involves projecting the point cloud into camera space and performing splatting operations to generate the final image. To blend color attributes accurately, points are sorted by depth, and pairwise compositing is applied to account for transparency. The compositing formula is:
\begin{equation}
C = \sum_{i \in N} c_i \prod_{j=1}^{i-1} (1 - \alpha_j),
\end{equation}

where \( c_i \) represents the color of point \( i \), and \( \alpha_i \) is its transparency.

\subsubsection{SMPL and Scene Alignment}
The SMPL model~\cite{loper2015smpl} represents human motion using:
\begin{itemize}
    \item Pose parameters: \( \theta \in \mathbb{R}^{24 \times 3} \), rotations of 24 body joints in axis-angle representation.
    \item Shape parameters : \( \beta \in \mathbb{R}^{10} \), body shape encoded as a 10-dimensional vector.
\end{itemize}

The posed mesh \( \mathbf{v} \) is computed from the rest pose \( \mathbf{v}^{\text{rest}} \) using Linear Blend Skinning (LBS):
\[
\mathbf{v}_i = \sum_{j=1}^{m} w_{ij} \mathbf{T}_j \mathbf{T}_j^0 \mathbf{v}_i^{\text{rest}},
\]
where
\begin{itemize}
    \item \(m\) is the number of joints.
    \item \( w_{ij} \) are skinning weights, \( \sum_{j=1}^{m} w_{ij} = 1, w_{ij} \geq 0 \),
    \item \( \mathbf{T}_j^0 \) transforms from the rest pose to the local joint space,
    \item \( \mathbf{T}_j \) transforms from local joint to target pose.
\end{itemize}

The background point clouds obtained via COLMAP~\cite{schonberger2016structure,schonberger2016pixelwise}  by applying the mask~\cite{wu2019detectron2} for humans and the SMPL human parameters estimated by ROMP~\cite{sun2021monocular} are in different coordinate systems, resulting in a misalignment between the two coordinates. To achieve alignment, similar to Neuman~\cite{jiang2022neuman}, we first use ROMP to estimate the 3D joints of the human body and their corresponding 2D projections in the image. Based on this projection, we solve the Perspective-n-Point (PnP) problem to obtain the initial extrinsic camera parameters, \( [R, t] \). However, additional scale estimation is required due to the inherent scale ambiguity of the PnP solution, which is insensitive to absolute distances. We assume the SMPL human model is standing on the ground. Therefore, we address this issue by solving for a scale factor \( s \) that aligns the SMPL model to the ground plane of the scene. The ground plane is fitted from the COLMAP scene point clouds and defined as 
\(
    ax + by + cz + d = 0,
\) where \( a, b, c, d \) are the coefficients of the ground plane. We set $\mathbf{A}$\(=(a,b,c) \).

Given the camera center $\mathbf{C}$\(=(c_x, c_y, c_z) \) and a 3D joint position $\mathbf{J}$\(=(j_x, j_y, j_z) \), the ray from the camera center to the joint can be expressed as:
\begin{equation}
    R(s) = \mathbf{C} + s \cdot (\mathbf{J}-\mathbf{C}),
\end{equation}
where \( s \) is the scale factor. Substituting this ray equation into the plane equation gives:
\begin{equation}
    s = -\frac{\mathbf{A}\cdot\mathbf{C} + d}{\mathbf{A}\cdot(\mathbf{J-C})}.
\end{equation}
By computing \( s \) for all joints, we select the minimum \( s \) to determine the appropriate scale factor for the SMPL model.

Using the calculated extrinsic camera parameters \( [R, t] \) and the scale factor \( s \), we achieve alignment between the human model and the scene.

\subsection{Method Overview}

Our approach retains the decoupled representation of background and human Gaussian point clouds, introducing a point cloud growing strategy based on a position texture to achieve richer human representation and enhanced detail capture. Additionally, we utilize a texture-driven PoP convolution network structure to predict the features of 3D Gaussian point clouds, making the point cloud representation more efficient. This enables our method to represent humans with fewer point clouds, resulting in significantly faster rendering speeds.

\begin{figure*}[h]
  \centering
  \includegraphics[width=\linewidth]{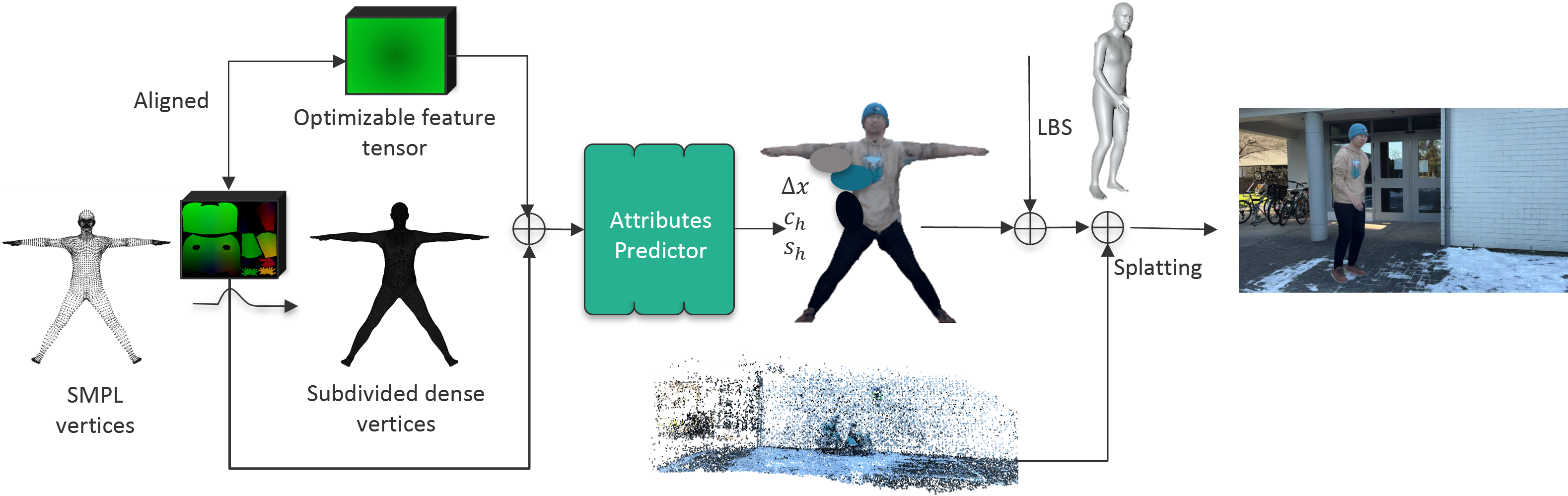}
  \caption{Overview of our SkinningGS framework. Starting from the SMPL model, the pipeline generates densely sampled vertices, predicts detailed attributes using an attribute predictor, and aligns them with the scene point cloud for rendering using splatting.}
  \label{fig:framework}
\end{figure*}

As shown in Figure~\ref{fig:framework}, our method framework begins with the SMPL model. To enhance the representation of the point cloud, we generate a dense set of points and their corresponding LBS weights using a position texture. These points are then processed and aligned with the scene's background point cloud obtained from COLMAP. 

The key step involves optimizing a feature tensor through the PoP predictor. This predictor is designed to capture the detailed attributes of each vertex, including its geometry offset \(\Delta x\), color $\mathbf{c}_h$, and scale $\mathbf{s}_h$. By leveraging the attribute predictor, we estimate how each vertex should deform, ensuring an accurate representation of human body geometry and appearance. 

Finally, we utilize the LBS  method to transform the SMPL model into the world coordinate and integrate it seamlessly with the background point cloud through splatting rendering. We optimize the human feature predictor and background points during this optimization process.

\subsection{Position Texture}

HUGS suffers from texture blurring and reliance on hyperparameter tuning of point densification. To address this problem, we introduce the position texture, which enables the generation of dense point clouds and maintains consistency during feature prediction and LBS weights calculation. It is represented as a tensor \(\boldsymbol{\tau} \in \mathbb{R}^{H \times W \times 3}\), where each pixel stores the 3D position of SMPL model vertices mapped onto a UV texture space. Beyond generating dense point clouds, the position texture facilitates efficient feature prediction using a PoP network, a convolutional neural network (CNN). The regular grid structure of the UV texture enables CNNs to predict point cloud features more effectively. In contrast to Gaussian Avatar~\cite{hu2024gaussianavatar}, which uses a position texture to record the pose information and optimize the pose feature, we simplify it only for points generation, which can accelerate the training speed and provide the chance for background joint-optimization with less parameters training.

The mapping process for the position texture is defined as:
\[
\mathbf{M} : 
\mathbf{p}_i \xrightarrow{\text{UV Mapping}} \mathbf{u}_i \xrightarrow{\text{Texture Sampling}} \boldsymbol{\tau}(\mathbf{u}_i),
\]
where \(\mathbf{p}_i \in \mathbb{R}^3\) represents the vertex position in 3D space, \(\mathbf{u}_i \in \mathbb{R}^2\) is the UV coordinate in the texture space, and \(\boldsymbol{\tau}(\mathbf{u}_i)\) denotes the interpolated position at \(\mathbf{u}_i\).

Each vertex \(\mathbf{p}_i\) of the SMPL model is mapped to the UV texture using a predefined UV mapping function~\cite{loper2015smpl} :
\[
\mathbf{u}_i = U(\mathbf{p}_i), \quad U : \mathbb{R}^3 \to \mathbb{R}^2,
\]
where \(U\) maps 3D positions to 2D UV coordinates. Additional points are interpolated within each triangle of the SMPL mesh. The 3D position for a point \(\mathbf{u} \in \mathbb{R}^2\) within a triangle is computed using barycentric interpolation:
\[
\boldsymbol{\tau}(\mathbf{u}) = \text{Interp}(U(\mathbf{p}_a), U(\mathbf{p}_b), U(\mathbf{p}_c); \mathbf{u}),
\]
where \(\mathbf{p}_a, \mathbf{p}_b, \mathbf{p}_c \in \mathbb{R}^3\) are 3D positions of the mapped vertices from SMPL. Similarly, LBS weights for the interpolated points are calculated using barycentric interpolation.

\begin{figure}[b!]
  \centering
  \includegraphics[width=\linewidth]{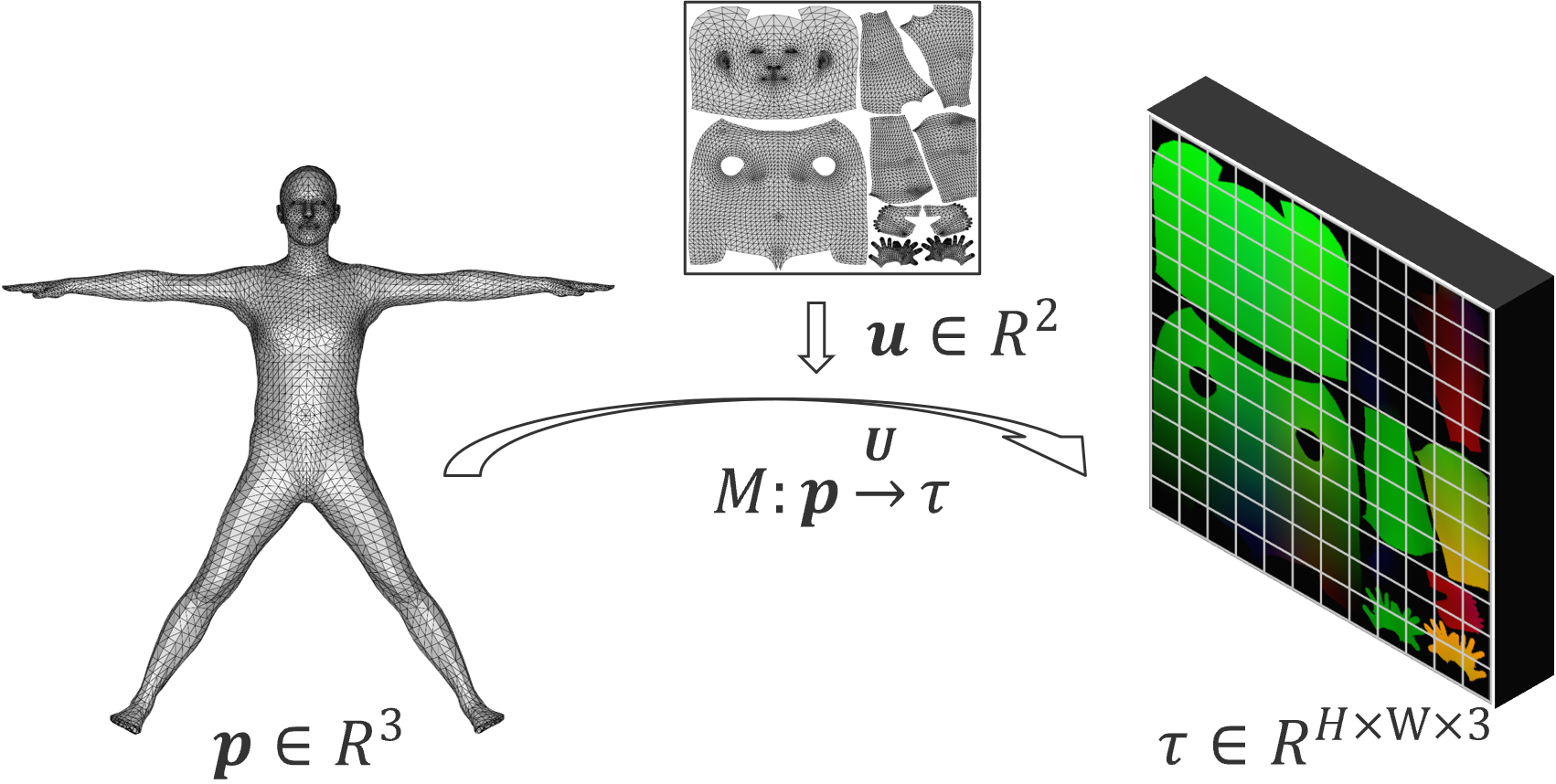}
  \caption{Position texture mapping of the SMPL mesh.}
  \label{fig:smpl_texture}
\end{figure}

We obtain the valid 3D position from position texture \(\boldsymbol{\tau}\) based on the 2D UV texture and initialize the human Gaussians using these 3D positions. The overall process, illustrated in Figure~\ref{fig:smpl_texture}, ensures high coverage, accurate geometry representation, and efficient feature prediction for downstream tasks.

\vspace{-5mm}
\subsection{Appearance-Forward Process}
In this approach, the Gaussian point is represented by several attributes: position \(\mathbf{p}\), opacity \(\boldsymbol{\alpha}\), RGB color \(\mathbf{c}\), and Gaussian kernel properties, including rotation \(r\) and scaling \(s\). For the background point cloud, these parameters are grouped into a set \( \mathcal{B}_\theta = \{\mathbf{p}_b, \boldsymbol{\alpha}_b, \mathbf{c}_b, r_b, s_b\}\) and we define \( \mathcal{H}_\theta = \{\mathbf{p}_h, \boldsymbol{\alpha}_h,\mathbf{c}_h, s_h\} \) for human point clouds. Our method retains the original characteristics of Gaussian Splatting for background points by densifying the point cloud to optimize rendering performance. However, for the human point cloud, the dynamic and potentially inaccurate pose and appearance make direct optimization suboptimal. A PoP structure network is introduced to generate refined features to address this challenge. We introduce an optimizable feature tensor \(\mathbf{F} \in \mathbb{R}^{H \times W \times D}\) aligned with the position texture, encoding the geometry features of human points in the canonical pose instead of the transformed pose~\cite{hu2024gaussianavatar} to \(D\) dimensions. This modification results in better properties prediction of geometry offset $\Delta x$, color $\mathbf{c_h}$, and scale $\mathbf{s_h}$ in this context.

The SMPL human model is initially T-posed in object space~\cite{loper2015smpl}. To reduce the overlap of the legs, we initialize the human body in canonical space as Da-Pose (looks like a Chinese character pronounced Da). The canonical pose is dynamically converted to the world coordinate system for each frame using the LBS. The transformation is expressed as follows:
\[
\mathbf{p}_{i,h} = \sum_{j=1}^{m} w_{ij} \mathbf{T}_j^{DW} \mathbf{T}_j^{TD} \left( \mathbf{v}_{i,h}^{\text{rest}} + \Delta x \right),
\]
where,
\begin{itemize}
    \item $\mathbf{T}_j^{TD}$ denotes the transformation from T-pose to Da-pose;
    \item $\mathbf{T}_j^{DW}$ represents the transformation from Da-pose to the expected pose in the world coordinate system.
\end{itemize}
The rendering process is defined as a Gaussian Splatting operation \( Splatting \), which takes both background and human parameters as input to obtain a rendered image:
\[
I = Splatting(\mathcal{B}_\theta, \mathcal{H}_\theta).
\]

\subsection{Training}
The optimization targets of our method include parameters for both the background and human point clouds. For the background point cloud, we optimize the parameters \( \mathcal{B}_\theta\), leveraging gradient-based updates in line with the 3DGS pipeline to ensure accurate scene geometry representation. 
For the human point cloud, the optimization focuses on LBS weights, opacity $\boldsymbol{\alpha}_h$, parameters of the point cloud feature prediction network, and an optimizable feature tensor. Human regions are optimized independently using masks to maintain proper separation from the background and enhance the human reconstruction quality.

The loss function is designed to achieve accurate reconstruction and effectively decoupling human and background components. It integrates multiple terms such as L1 loss, SSIM loss~\cite{wang2004image}, and LPIPS loss~\cite{zhang2018unreasonable} for reconstruction quality and perceptual consistency. Additionally, L2 penalties are applied to geometry tensors, predicted offsets, and scales to reduce noise and stabilize the point cloud features.

The final loss function is formulated as follows:
\begin{multline*}
\mathcal{L} = 
\overbrace{\lambda_1 \mathcal{L}_1 + \lambda_2 \mathcal{L}_{\text{SSIM}} + \lambda_3 \mathcal{L}_{\text{LPIPS}}}^{\text{scene + human}}
+ \underbrace{\lambda_1 \mathcal{L}_1^h + \lambda_2 \mathcal{L}_{\text{SSIM}}^h + \lambda_3 \mathcal{L}_{\text{LPIPS}}^h}_{\text{human}} \\
+ \lambda_4 \mathcal{L}_{\text{geo}}
+ \lambda_5 \mathcal{L}_{\text{offset}}
+ \lambda_6 \mathcal{L}_{\text{scale}}
\end{multline*}

where \(\lambda_1\), \(\lambda_2\), and \(\lambda_3\) contribute to both the human and scene reconstruction.
The background optimization employs a densification scheme to increase the point density incrementally. In our experiments, we set the following values for the loss weights:
\[
\lambda_1 = 0.7, \quad \lambda_2 = 0.3, \quad \lambda_3 = \lambda_4=\lambda_5=\lambda_6 = 1.0.
\]

\section{Experiments}

\subsection{Datasets}
\textbf{Neuman Dataset} consists of six video sequences, primarily focusing on humans performing actions in outdoor environments and highlighting diverse interactions between humans and scenes.

\textbf{HumanCap Dataset} ~\cite{vlasic2008articulated,xu2018monoperfcap,perazzi2016benchmark,yang2023reconstructing} contains various human videos, and we select four test sequences. It captures simple human motion in indoor environments with a single camera. This dataset emphasizes scenarios with fewer human-scene interactions, enabling isolated evaluation of human motion.

\subsection{Reconstruction Quality Metrics}
We compared our method against several baselines, including HUGS and the NeRF-based approach Neuman. For focused human-scene contacted area analysis, we employed a mask for humans, cropping the human and its surrounding regions from the bounding box.

We use PSNR, SSIM~\cite{wang2004image}, and LPIPS~\cite{zhang2018unreasonable} as metrics for evaluation. The results are presented in Tables~\ref{tab:neuman_comparison_1} and~\ref{tab:neuman_comparison_2}, covering six datasets: \textit{bike}, \textit{citron}, \textit{jogging}, \textit{lab}, \textit{parkinglot}, and \textit{seattle}. As you can see from the Tables~\ref{tab:neuman_comparison_1} and~\ref{tab:neuman_comparison_2}, all of our reconstruction metrics can outperform HUGS, and the quality of our reconstruction has improved except for a slightly worse performance on the citron dataset. In all tables, \textcolor{ForestGreen}{green} indicates the best and \textcolor{Goldenrod}{yellow} indicates a suboptimal result.

\begin{table*}[t]
\centering
\resizebox{\textwidth}{!}{%
\renewcommand{\arraystretch}{1.2}
\begin{tabular}{c cccccc cccccc cccccc}
\hline
 \multirow[c]{3}{*}{Method}  & \multicolumn{6}{c}{{\textbf{Bike}}} & \multicolumn{6}{c}{{\textbf{Citron}}} & \multicolumn{6}{c}{{\textbf{Jogging}}} \\ 
 \cmidrule(lr){2-7} \cmidrule(lr){8-13} \cmidrule(lr){14-19}
 & \multicolumn{3}{c}{{SceneHuman}}&\multicolumn{3}{c}{{Human}} & \multicolumn{3}{c}{{SceneHuman}} &\multicolumn{3}{c}{{Human}}& \multicolumn{3}{c}{{SceneHuman}}&\multicolumn{3}{c}{{Human}} \\
 &\multicolumn{3}{c}{{ PSNR↑  SSIM↑  LPIPS↓}}&\multicolumn{3}{c}{{ PSNR↑  SSIM↑  LPIPS↓}}&\multicolumn{3}{c}{{ PSNR↑  SSIM↑  LPIPS↓}}&\multicolumn{3}{c}{{ PSNR↑  SSIM↑  LPIPS↓}}&\multicolumn{3}{c}{{ PSNR↑  SSIM↑  LPIPS↓}}&\multicolumn{3}{c}{{ PSNR↑  SSIM↑  LPIPS↓}} \\ \hline
Neuman & \cellcolor{yellow!75}25.554 & 0.830 & 0.166 & 19.102 & 0.662 & 0.173 & 24.774 & 0.811 & 0.178 & 18.691 & 0.639 & 0.161 & 22.697 & 0.681 & 0.273 & \cellcolor{yellow!75}17.572 & 0.538 & 0.274 \\ \hline
HUGS   & 25.457 & \cellcolor{yellow!75}0.844 &\cellcolor{yellow!75} 0.097 & \cellcolor{yellow!75}19.477 & \cellcolor{yellow!75}0.674 & \cellcolor{yellow!75}0.158 & \cellcolor{ForestGreen!50}25.541 & \cellcolor{ForestGreen!50}0.860 & \cellcolor{ForestGreen!50}0.095 &\cellcolor{yellow!75} 19.157 & \cellcolor{ForestGreen!50}0.707 & \cellcolor{ForestGreen!50}0.133 & \cellcolor{yellow!75}23.747 & \cellcolor{yellow!75}0.779 &\cellcolor{yellow!75} 0.177 & 17.452 &\cellcolor{yellow!75} 0.589 & \cellcolor{yellow!75}0.256 \\ \hline
\textbf{Ours}   &\cellcolor{ForestGreen!50} 26.601 & \cellcolor{ForestGreen!50}0.876 &\cellcolor{ForestGreen!50} 0.080 &\cellcolor{ForestGreen!50} 20.391 & \cellcolor{ForestGreen!50}0.709 &\cellcolor{ForestGreen!50} 0.137 & \cellcolor{yellow!75}25.539 &\cellcolor{yellow!75} 0.846 & \cellcolor{yellow!75}0.125 & \cellcolor{ForestGreen!50}19.457 &\cellcolor{yellow!75} 0.694 & \cellcolor{yellow!75}0.139 &\cellcolor{ForestGreen!50} 24.288 &\cellcolor{ForestGreen!50} 0.809 &\cellcolor{ForestGreen!50} 0.164 & \cellcolor{ForestGreen!50}18.001 & \cellcolor{ForestGreen!50}0.599 & \cellcolor{ForestGreen!50}0.223 \\ \hline
\end{tabular}%
}
\caption{Comparison of evaluation metrics across different methods for the Neuman bike, citron, and jogging samples dataset.}
\label{tab:neuman_comparison_1}
\end{table*}

\begin{table*}[t]
\centering
\resizebox{\textwidth}{!}{%
\renewcommand{\arraystretch}{1.2}
\begin{tabular}{c cccccc cccccc cccccc}
\hline
 \multirow{3}{*}{Method} & \multicolumn{6}{c}{{\textbf{Lab}}} & \multicolumn{6}{c}{{\textbf{Parkinglot}}} & \multicolumn{6}{c}{{\textbf{Seattle}}} \\ 
 \cmidrule(lr){2-7} \cmidrule(lr){8-13} \cmidrule(lr){14-19}
 & \multicolumn{3}{c}{{SceneHuman}}&\multicolumn{3}{c}{{Human}} & \multicolumn{3}{c}{{SceneHuman}} &\multicolumn{3}{c}{{Human}}& \multicolumn{3}{c}{{SceneHuman}}&\multicolumn{3}{c}{{Human}} \\
 &\multicolumn{3}{c}{{ PSNR↑  SSIM↑  LPIPS↓}}&\multicolumn{3}{c}{{ PSNR↑  SSIM↑  LPIPS↓}}&\multicolumn{3}{c}{{ PSNR↑  SSIM↑  LPIPS↓}}&\multicolumn{3}{c}{{ PSNR↑  SSIM↑  LPIPS↓}}&\multicolumn{3}{c}{{ PSNR↑  SSIM↑  LPIPS↓}}&\multicolumn{3}{c}{{ PSNR↑  SSIM↑  LPIPS↓}} \\ \hline
Neuman & 24.960 & 0.862 & 0.149 & 18.756 & 0.726 & 0.193 & 25.434 & 0.800 & 0.201 & 17.663 & 0.660 & 0.205 & 23.986 & 0.782 & 0.194 & 18.416 & 0.578 & 0.186 \\ \hline
HUGS   & \cellcolor{yellow!75}25.997 & \cellcolor{yellow!75}0.916 & \cellcolor{ForestGreen!50}0.070 & \cellcolor{yellow!75}18.790 & \cellcolor{yellow!75}0.762 & \cellcolor{yellow!75}0.152 & \cellcolor{yellow!75}26.862 & \cellcolor{yellow!75}0.850 & \cellcolor{yellow!75}0.136 & \cellcolor{yellow!75}19.437 & \cellcolor{yellow!75}0.730 & \cellcolor{yellow!75}0.151 & \cellcolor{yellow!75}25.937 & \cellcolor{yellow!75}0.853 & \cellcolor{yellow!75}0.093 & \cellcolor{yellow!75}19.061 & \cellcolor{yellow!75}0.668 & \cellcolor{yellow!75}0.141 \\ \hline
\textbf{Ours}   & \cellcolor{ForestGreen!50}26.378 & \cellcolor{ForestGreen!50}0.917 & \cellcolor{yellow!75}0.071 & \cellcolor{ForestGreen!50}19.491 & \cellcolor{ForestGreen!50}0.766 & \cellcolor{ForestGreen!50}0.147 & \cellcolor{ForestGreen!50}27.947 & \cellcolor{ForestGreen!50}0.861 & \cellcolor{ForestGreen!50}0.129 & \cellcolor{ForestGreen!50}20.704 & \cellcolor{ForestGreen!50}0.743 & \cellcolor{ForestGreen!50}0.135 & \cellcolor{ForestGreen!50}27.267 & \cellcolor{ForestGreen!50}0.896 & \cellcolor{ForestGreen!50}0.081 & \cellcolor{ForestGreen!50}20.314 & \cellcolor{ForestGreen!50}0.712 & \cellcolor{ForestGreen!50}0.135 \\ \hline

\end{tabular}%
}
\caption{Comparison of evaluation metrics across different methods for the Neuman lab, parkinglot, and seattle samples dataset.}
\label{tab:neuman_comparison_2}
\end{table*}


\begin{table*}[t]
\centering
\resizebox{\textwidth}{!}{%
\renewcommand{\arraystretch}{1.4}
\begin{tabular}{c cccccc cccccc cccccc cccccc}
\hline
 \multirow{3}{*}{Method} & \multicolumn{6}{c}{{\textbf{HumanCap\_0}}} & \multicolumn{6}{c}{{\textbf{HumanCap\_2}}} & \multicolumn{6}{c}{{\textbf{HumanCap\_3}}} & \multicolumn{6}{c}{{\textbf{HumanCap\_9}}} \\ 
 \cmidrule(lr){2-7} \cmidrule(lr){8-13} \cmidrule(lr){14-19} \cmidrule(lr){20-25}
 & \multicolumn{3}{c}{{SceneHuman}} & \multicolumn{3}{c}{{Human}} & \multicolumn{3}{c}{{SceneHuman}} & \multicolumn{3}{c}{{Human}} & \multicolumn{3}{c}{{SceneHuman}} & \multicolumn{3}{c}{{Human}} & \multicolumn{3}{c}{{SceneHuman}} & \multicolumn{3}{c}{{Human}} \\
 & \multicolumn{3}{c}{{ PSNR↑  SSIM↑  LPIPS↓}} & \multicolumn{3}{c}{{ PSNR↑  SSIM↑  LPIPS↓}} & \multicolumn{3}{c}{{ PSNR↑  SSIM↑  LPIPS↓}} & \multicolumn{3}{c}{{ PSNR↑  SSIM↑  LPIPS↓}} & \multicolumn{3}{c}{{ PSNR↑  SSIM↑  LPIPS↓}} & \multicolumn{3}{c}{{ PSNR↑  SSIM↑  LPIPS↓}} & \multicolumn{3}{c}{{ PSNR↑  SSIM↑  LPIPS↓}} & \multicolumn{3}{c}{{ PSNR↑  SSIM↑  LPIPS↓}} \\ \hline
HUGS   & 25.425 & 0.886 & 0.081 & 21.168 & 0.760 & 0.155 & 26.193 & 0.901 & 0.069 & 21.579 & 0.762 & 0.150 & 25.120 & 0.877 & 0.119 & 21.629 & 0.811 & 0.141 & 22.118 & 0.790 & 0.205 & 17.336 & 0.634 & 0.330 \\ \hline
\textbf{Ours}   & \cellcolor{ForestGreen!50}26.461 & \cellcolor{ForestGreen!50}0.903 & \cellcolor{ForestGreen!50}0.072 & \cellcolor{ForestGreen!50}22.331 & \cellcolor{ForestGreen!50}0.792 & \cellcolor{ForestGreen!50}0.141 & \cellcolor{ForestGreen!50}26.461 & \cellcolor{ForestGreen!50}0.903 & \cellcolor{ForestGreen!50}0.072 & \cellcolor{ForestGreen!50}22.331 & \cellcolor{ForestGreen!50}0.792 & \cellcolor{ForestGreen!50}0.141 & \cellcolor{ForestGreen!50}26.073 & \cellcolor{ForestGreen!50}0.890 & \cellcolor{ForestGreen!50}0.108 & \cellcolor{ForestGreen!50}22.597 & \cellcolor{ForestGreen!50}0.828 & \cellcolor{ForestGreen!50}0.140 & \cellcolor{ForestGreen!50}22.752 & \cellcolor{ForestGreen!50}0.843 & \cellcolor{ForestGreen!50}0.153 & \cellcolor{ForestGreen!50}17.706 & \cellcolor{ForestGreen!50}0.654 & \cellcolor{ForestGreen!50}0.326 \\ \hline
\end{tabular}%
}
\caption{Comparison of evaluation metrics across HumamCap samples.}
\label{tab:comparison_human_cap}
\end{table*}

\begin{table*}[h!]
\centering
\Large
\resizebox{\textwidth}{!}{%
\setlength{\tabcolsep}{5pt}
\renewcommand{\arraystretch}{1.1}
\begin{tabular}{c cccccc cccccc cccccc}
\hline
\multirow{3}{*}{Method} & 
\multicolumn{6}{c}{\textbf{Bike}} & 
\multicolumn{6}{c}{\textbf{Citron}} & 
\multicolumn{6}{c}{\textbf{Jogging}} \\
\cmidrule(lr){2-7} \cmidrule(lr){8-13} \cmidrule(lr){14-19}
& \multicolumn{3}{c}{SceneHuman} & \multicolumn{3}{c}{Human} & 
\multicolumn{3}{c}{SceneHuman} & \multicolumn{3}{c}{Human} & 
\multicolumn{3}{c}{SceneHuman} & \multicolumn{3}{c}{Human} \\
& PSNR↑ & SSIM↑ & LPIPS↓ & PSNR↑ & SSIM↑ & LPIPS↓ & 
PSNR↑ & SSIM↑ & LPIPS↓ & PSNR↑ & SSIM↑ & LPIPS↓ & 
PSNR↑ & SSIM↑ & LPIPS↓ & PSNR↑ & SSIM↑ & LPIPS↓ \\ \hline
Baseline & 26.412 & \cellcolor{ForestGreen!50}0.877 & \cellcolor{ForestGreen!50}0.0799 & 20.090 & 0.706 & 0.144 & \cellcolor{yellow!75}25.841 & \cellcolor{ForestGreen!50}0.865 & \cellcolor{ForestGreen!50}0.093 & 19.208 & 0.694 & 0.147 & 24.172 & \cellcolor{yellow!75}0.810 & \cellcolor{yellow!75}0.164 & 17.792 & \cellcolor{ForestGreen!50}0.600 & 0.237 \\ 
w/o LBS & 26.476 & 0.875 & \cellcolor{yellow!75}0.080 & 20.244 & 0.708 & \cellcolor{yellow!75}0.140 &  \cellcolor{ForestGreen!50}26.051 & \cellcolor{yellow!75}0.864 & \cellcolor{ForestGreen!50}0.093 &\cellcolor{ForestGreen!50} 19.522 & \cellcolor{yellow!75}0.696 & \cellcolor{ForestGreen!50}0.135 & 24.059 & 0.807 & 0.166 & 17.551 & 0.591 & \cellcolor{yellow!75}0.227 \\ 
w/o opacities & \cellcolor{yellow!75}26.510 & 0.874 & 0.081 & \cellcolor{ForestGreen!50}20.393 & \cellcolor{yellow!75}0.704 & \cellcolor{yellow!75}0.140 & 25.808 & 0.863 & \cellcolor{yellow!75}0.095 & 19.383 & \cellcolor{ForestGreen!50}0.697 & 0.147 & \cellcolor{yellow!75}24.272 & \cellcolor{ForestGreen!50}0.813 & \cellcolor{ForestGreen!50}0.163 &\cellcolor{ForestGreen!50} 18.107 & \cellcolor{yellow!75}0.599 & 0.238 \\ 
Ours & \cellcolor{ForestGreen!50}26.601 & \cellcolor{yellow!75}0.876 & \cellcolor{yellow!75}0.080 & \cellcolor{yellow!75}20.391 & \cellcolor{ForestGreen!50}0.709 & \cellcolor{ForestGreen!50}0.137 & 25.539 & 0.846 & 0.125 & \cellcolor{yellow!75}19.457 & 0.694 &\cellcolor{yellow!75} 0.139 & \cellcolor{ForestGreen!50}24.288 & 0.809 & \cellcolor{yellow!75}0.164 & \cellcolor{yellow!75}18.001 & \cellcolor{yellow!75}0.599 & \cellcolor{ForestGreen!50}0.223 \\ \hline
\end{tabular}%
}
\caption{Ablation study metrics on Bike, Citron, Jogging scenes.}
\label{tab:ablation_metrics_part1}
\end{table*}
\begin{table*}[h!]
\centering
\Large
\resizebox{\textwidth}{!}{%
\setlength{\tabcolsep}{5pt}
\renewcommand{\arraystretch}{1.2}
\begin{tabular}{c cccccc cccccc cccccc}
\hline
\multirow{3}{*}{Method} & 
\multicolumn{6}{c}{\textbf{Lab}} & 
\multicolumn{6}{c}{\textbf{Parkinglot}} & 
\multicolumn{6}{c}{\textbf{Seattle}} \\
\cmidrule(lr){2-7} \cmidrule(lr){8-13} \cmidrule(lr){14-19}
& \multicolumn{3}{c}{SceneHuman} & \multicolumn{3}{c}{Human} & 
\multicolumn{3}{c}{SceneHuman} & \multicolumn{3}{c}{Human} & 
\multicolumn{3}{c}{SceneHuman} & \multicolumn{3}{c}{Human} \\
& PSNR↑ & SSIM↑ & LPIPS↓ & PSNR↑ & SSIM↑ & LPIPS↓ & 
PSNR↑ & SSIM↑ & LPIPS↓ & PSNR↑ & SSIM↑ & LPIPS↓ & 
PSNR↑ & SSIM↑ & LPIPS↓ & PSNR↑ & SSIM↑ & LPIPS↓ \\ \hline
Baseline & 26.249 & \cellcolor{yellow!75}0.916 & 0.075 & 19.396 & 0.765 & 0.153 & 27.149 & 0.855 & 0.137 & 19.823 & 0.731 & 0.150 & 27.053 & 0.894 & \cellcolor{ForestGreen!50}0.081 & 19.378 & 0.693 & 0.143 \\ 
w/o LBS & 26.337 & \cellcolor{ForestGreen!50}0.917 & \cellcolor{ForestGreen!50}0.071 & \cellcolor{yellow!75}19.479 & \cellcolor{yellow!75}0.766 & \cellcolor{ForestGreen!50}0.147 & 27.465 & \cellcolor{yellow!75}0.858 & \cellcolor{yellow!75}0.132 & 20.168 & 0.732 & 0.144 & 27.086 & 0.894 & 0.082 & 19.461 & 0.697 & \cellcolor{yellow!75}0.141 \\ 
w/o opacities & \cellcolor{ForestGreen!50}26.391 & \cellcolor{yellow!75}0.916 & 0.075 & 19.471 & \cellcolor{ForestGreen!50}0.768 & \cellcolor{yellow!75}0.152 & \cellcolor{yellow!75}27.728 & 0.853 & 0.134 & \cellcolor{yellow!75}20.597 & \cellcolor{yellow!75}0.738 & \cellcolor{yellow!75}0.138 &\cellcolor{ForestGreen!50} 27.397 & \cellcolor{yellow!75}0.895 & \cellcolor{ForestGreen!50}0.081 & \cellcolor{yellow!75}20.085 & \cellcolor{yellow!75}0.706 & \cellcolor{yellow!75}0.141 \\ 
Ours & \cellcolor{yellow!75}26.378 & \cellcolor{ForestGreen!50}0.917 & \cellcolor{ForestGreen!50}0.071 & \cellcolor{ForestGreen!50}19.491 & \cellcolor{yellow!75}0.766 & \cellcolor{ForestGreen!50}0.147 & \cellcolor{ForestGreen!50}27.947 & \cellcolor{ForestGreen!50}0.861 & \cellcolor{ForestGreen!50}0.129 & \cellcolor{ForestGreen!50}20.704 & \cellcolor{ForestGreen!50}0.743 & \cellcolor{ForestGreen!50}0.135 & \cellcolor{yellow!75}27.267 & \cellcolor{ForestGreen!50}0.896 & \cellcolor{ForestGreen!50}0.081 & \cellcolor{ForestGreen!50}20.314 & \cellcolor{ForestGreen!50}0.712 & \cellcolor{ForestGreen!50}0.135 \\ \hline
\end{tabular}%
}
\caption{Ablation study metrics on Lab, Parkinglot, Seattle scenes.}
\label{tab:ablation_metrics_part2}
\end{table*}

Meanwhile, a comparison between our method and HUGS is shown in Table~\ref{tab:comparison_human_cap}, where the quality of our reconstruction exceeds that of HUGS in four samples of the HumanCap dataset.

\begin{figure}[!ht]
  \centering
  \includegraphics[width=\linewidth]{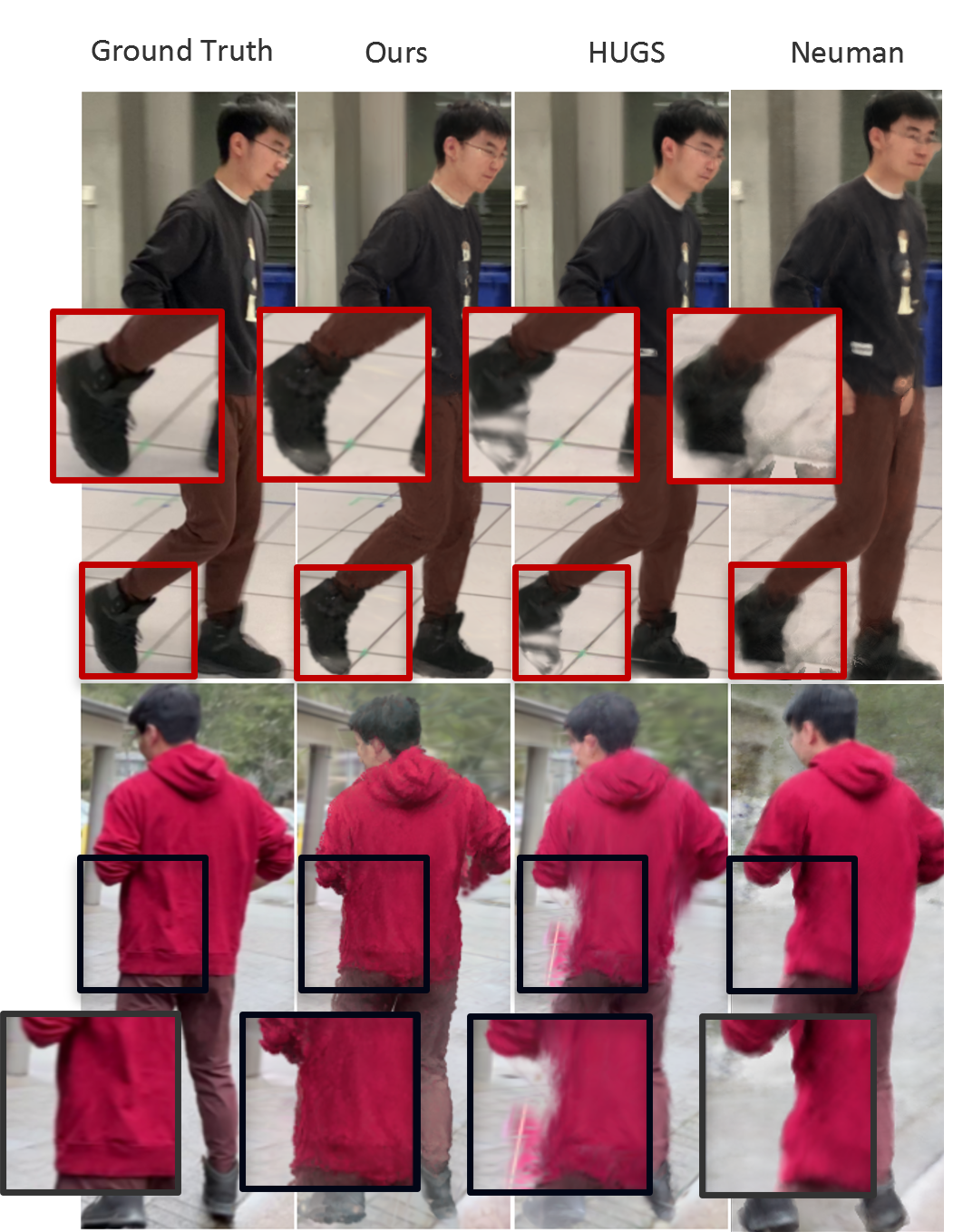}
  \caption{Comparison over Neuman dataset visually.}
  \label{fig:neuman_dataset}
\end{figure}

\begin{figure}[h]
  \centering
  \includegraphics[width=\linewidth]{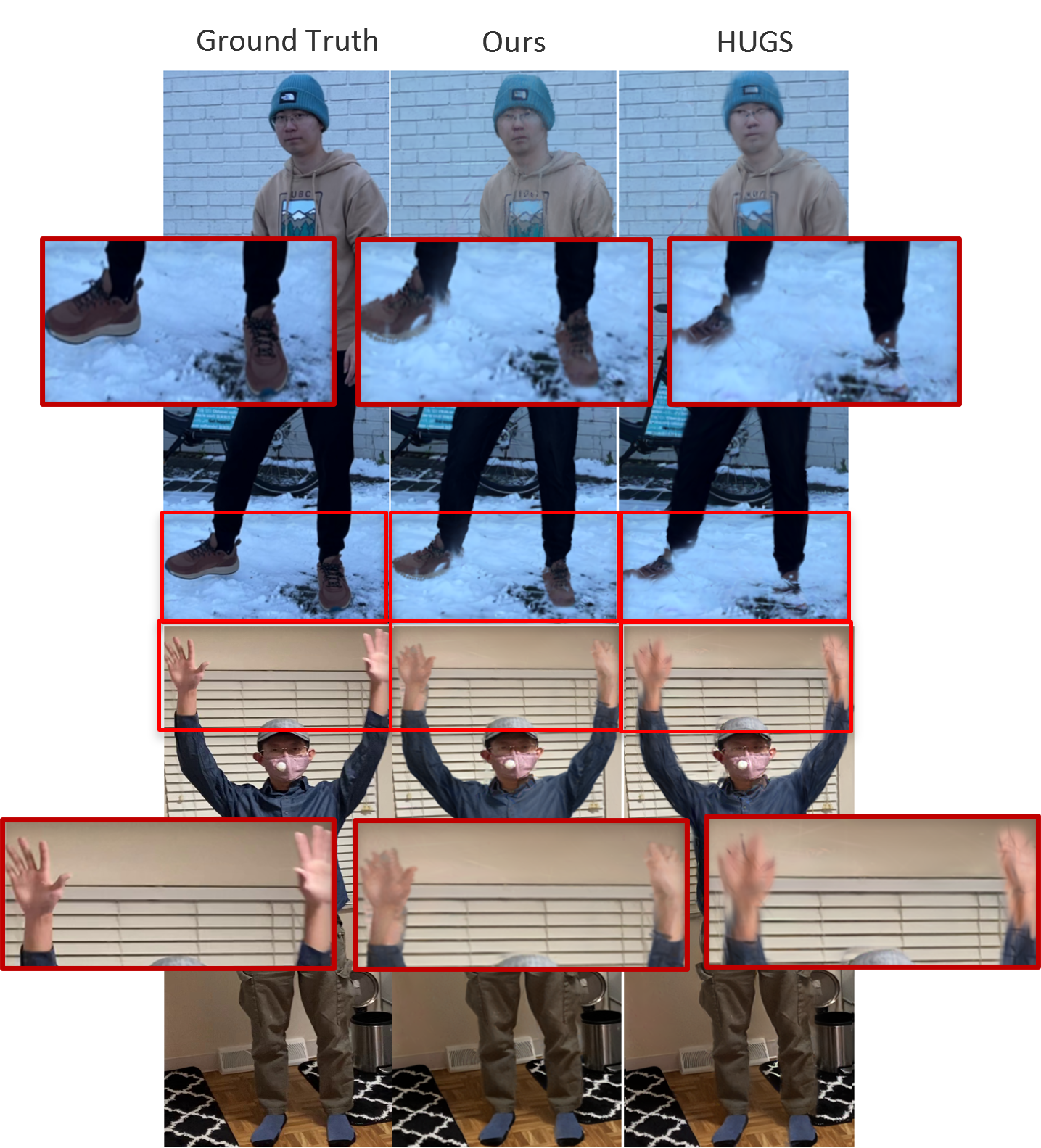}
  \caption{Comparison over Neuman and Humancap dataset visually.}
  \label{fig:humancap_dataset}
\end{figure}

We also highlight the visual differences in reconstruction quality achieved by various methods. In Figure~\ref{fig:neuman_dataset} and Figure~\ref{fig:humancap_dataset}, we observe a recurring issue in methods such as HUGS and Neuman when the human body interacts with the ground. Specifically, background point clouds often partially obstruct the feet, leading to an inaccurate representation of depth and transparency. In contrast, our approach precisely reconstructs the relationship between the human body and the background in terms of depth and transparency.

Furthermore, in the lower part of Figure~\ref{fig:neuman_dataset}, our method demonstrates the ability to reduce the occurrence of floaters—artifacts commonly seen in reconstructing clothing. This improvement ensures a more accurate representation of clothing geometry, which is crucial for high-quality visual outputs. Figure~\ref{fig:neuman_dataset} also shows that our approach outperforms HUGS in reconstructing fine details, delivering superior clarity and structural accuracy.

\begin{figure}[h]
  \centering
  \includegraphics[width=\linewidth]{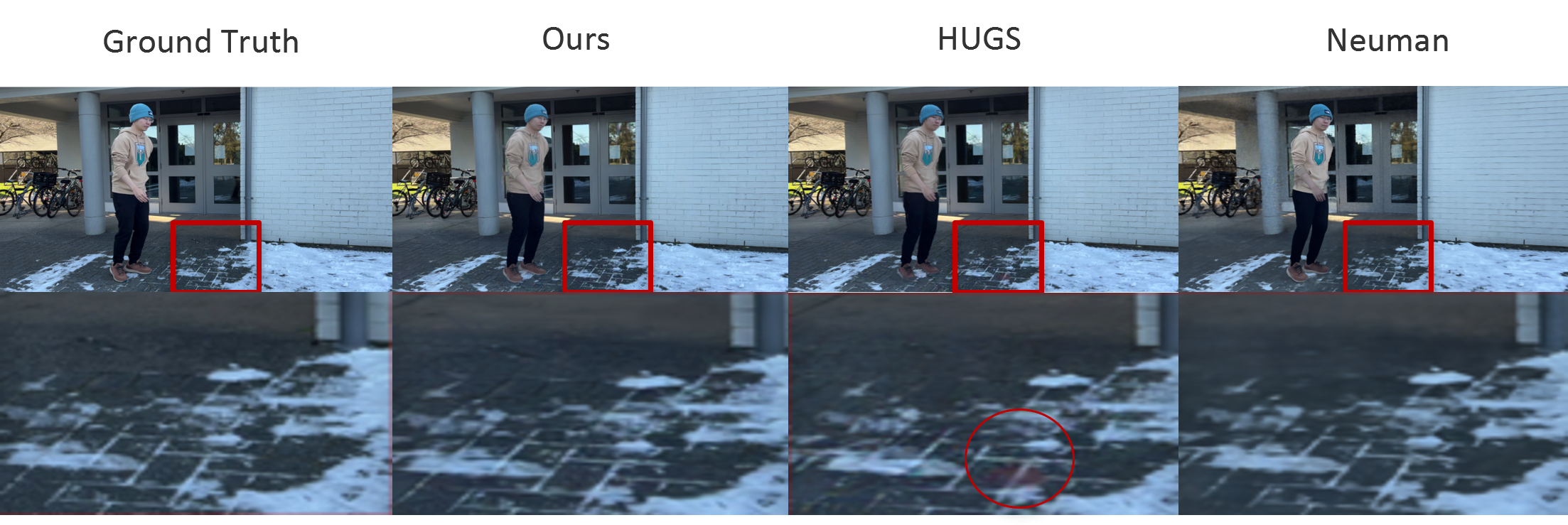}
  \caption{Comparison about the background reconstruction visually.}
  \label{fig:bg_comparison}
\end{figure}

Figure~\ref{fig:bg_comparison} showcases the results of joint optimization of the human body and the background. In the case of HUGS, which utilizes deformation fields for optimization, floaters often appear within the background, degrading the overall visual quality. In contrast, our method successfully eliminates such artifacts, resulting in a more cohesive and artifact-free reconstruction of the human body and the background.

\subsection{Efficient Human Representation and Fast Offline Rendering}
We compare our method with HUGS regarding human representation efficiency and offline rendering speed. Our experiments were conducted using NVIDIA RTX 4090 GPUs for training, with the training process typically taking 1–2 hours. Offline rendering speed was evaluated on NVIDIA RTX 4090 and NVIDIA RTX 3070 GPUs and calculated based on LBS transformation and rendering time.

Our method achieves efficient human representations by utilizing approximately 200k points—only half the number required by HUGS—while delivering superior performance. The reduced number of point clouds minimizes the need to calculate deformation field weights, accelerating the offline rendering process. As a result, our method achieves approximately 190 FPS on the NVIDIA RTX 4090, making it $\sim$5$\times$ faster than HUGS, and around 135 FPS on the NVIDIA RTX 3070, reaching $\sim$6$\times$ the speed of HUGS. These results highlight the significant advantage of our approach in fast and efficient rendering.




\subsection{Ablation Study}
We conducted an ablation study to investigate opacities and LBS weights on the reconstruction quality of our framework. This section demonstrates how these components influence human and scene reconstruction through visual comparisons and quantitative metrics.

\begin{figure}[h]
  \centering
  \includegraphics[width=\linewidth]{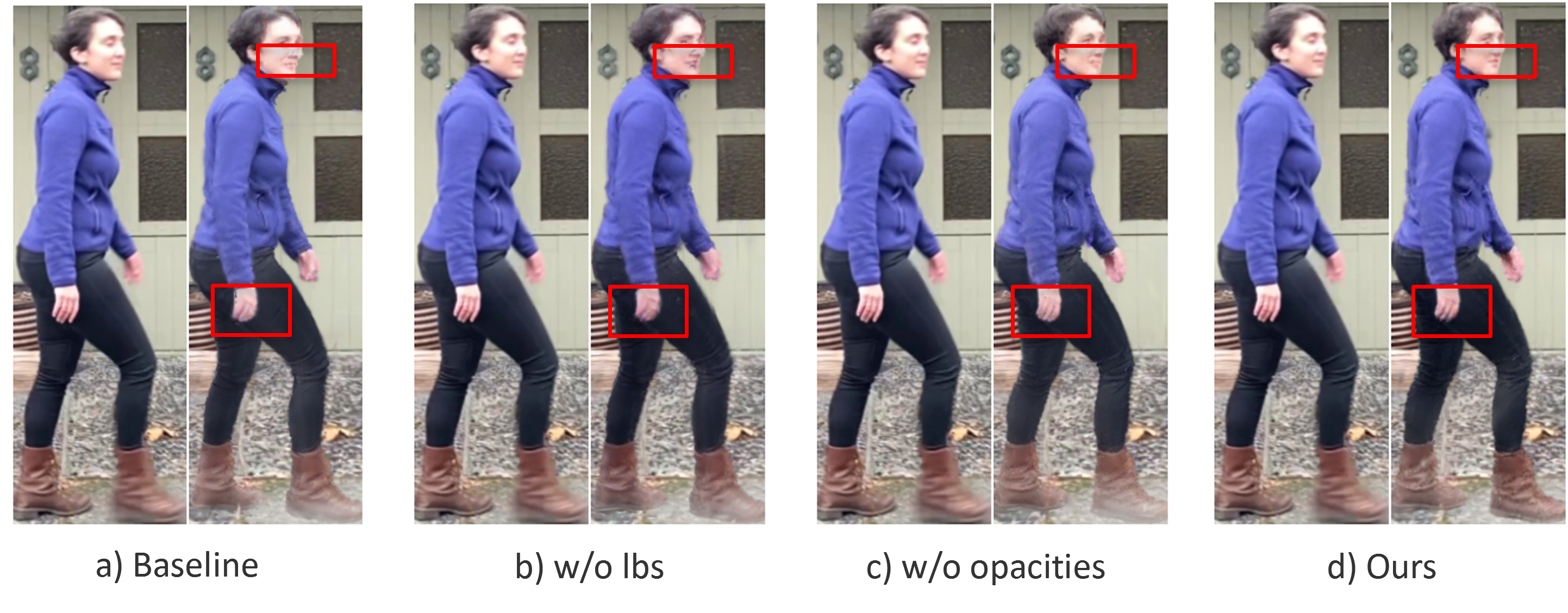}
  \caption{Comparison for ablation study visually.}
  \label{fig:ablation}
\end{figure}

Results show that incorporating either transparency optimization or LBS alone improves reconstruction but still leaves gaps in detail or poses accuracy. Combining both enhances overall performance, achieving better reconstruction quality. Figure~\ref{fig:ablation} demonstrates the visual differences under various configurations. In a) we can see hand textures fail to reconstruct detailed patterns, and pose accuracy suffers. In b) only transparency optimization might cause distortions in facial details. In c) hand details are better reconstructed. In d), by combining both optimizations, our method achieves superior results, accurately restoring fine facial textures and hand details. Tables~\ref{tab:ablation_metrics_part1} and~\ref{tab:ablation_metrics_part2} quantitatively compare PSNR, SSIM, and LPIPS metrics across multiple data samples in the Neuman dataset. The combined transparency and LBS optimization usually outperform other settings, particularly in "parkinglot" and "seattle" scenarios.

\subsection{Ablation for Different Resolution of Position Texture}
At the same time, we also conducted an ablation study to investigate the impact of position texture resolution on the reconstruction quality of our framework over the bike sample. 
Table~\ref{tab:resolution_ablation} shows that increasing the resolution of Position-Texture enhances reconstruction quality metrics. At the highest resolution (512×512), the PSNR improves to 26.0641, and the LPIPS decreases to 0.0988, indicating superior rendering fidelity. However, the training time grows significantly, increasing from 15 minutes at 128×128 to 45 minutes at 512×512. This highlights the trade-off between computational efficiency and rendering quality, providing critical insights for balancing these factors in practical applications.
\begin{table}[h!]
    \centering
    \renewcommand{\arraystretch}{2} 
    \setlength{\tabcolsep}{4pt} 
    \begin{tabular}{c c c c c}
        \hline
        \textbf{Resolution} & \textbf{Training Time} & \textbf{PSNR↑} & \textbf{SSIM↑} & \textbf{ LPIPS↓} \\
         \hline
        128×128 & 15 mins & 25.8432 & 0.8361 & 0.1284  \\ \hline
        256×256 & 25 mins & \cellcolor{yellow!75}25.9766 & \cellcolor{yellow!75}0.8502 & \cellcolor{yellow!75}0.1079  \\ \hline
        512×512 & 45 mins & \cellcolor{ForestGreen!50}26.601 & \cellcolor{ForestGreen!50}0.876 & \cellcolor{ForestGreen!50}0.080  \\ \hline
    \end{tabular}
    \caption{Comparison of performance metrics for different position texture resolutions. Training was done on NVIDIA RTX4090. All the metrics are evaluated for the whole image.}
    \label{tab:resolution_ablation}
\end{table}

\vspace{-10mm}
\section{Applications}
One of the significant advantages of separating the background and human point clouds is the ability to retarget human motion. This section demonstrates some related applications: real-time motion editing and novel motion synthesis. At the same time, we show the texture-based method can also easily be extended to animal-scene reconstruction.

\subsection{Real-Time Motion Editing}
\begin{figure}[h]
  \centering
  \includegraphics[width=\linewidth]{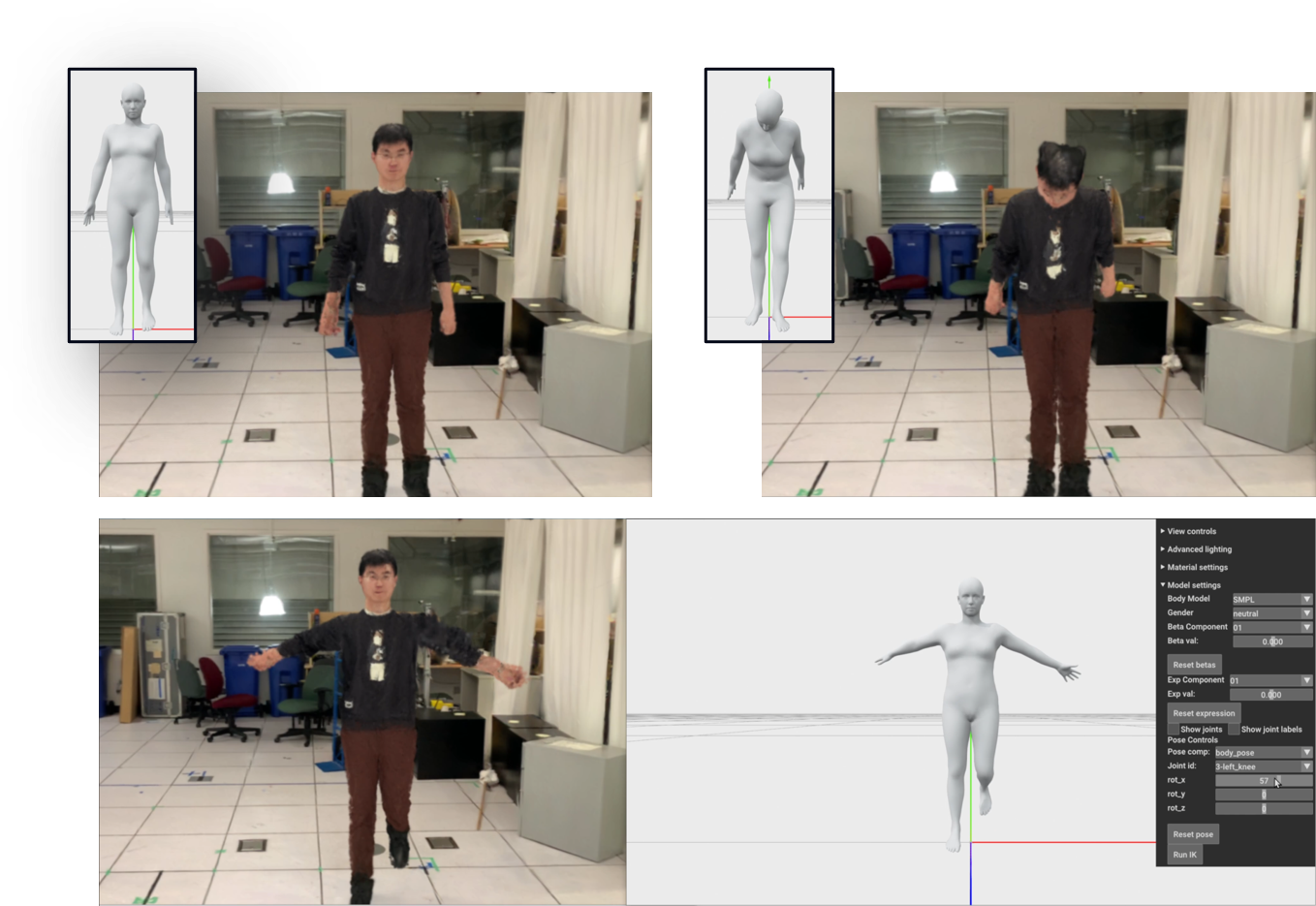}
  \caption{Motion editing in real-time.}
  \label{fig:motion_edit}
\end{figure}

\begin{figure}[h]
  \centering
  \includegraphics[width=\linewidth]{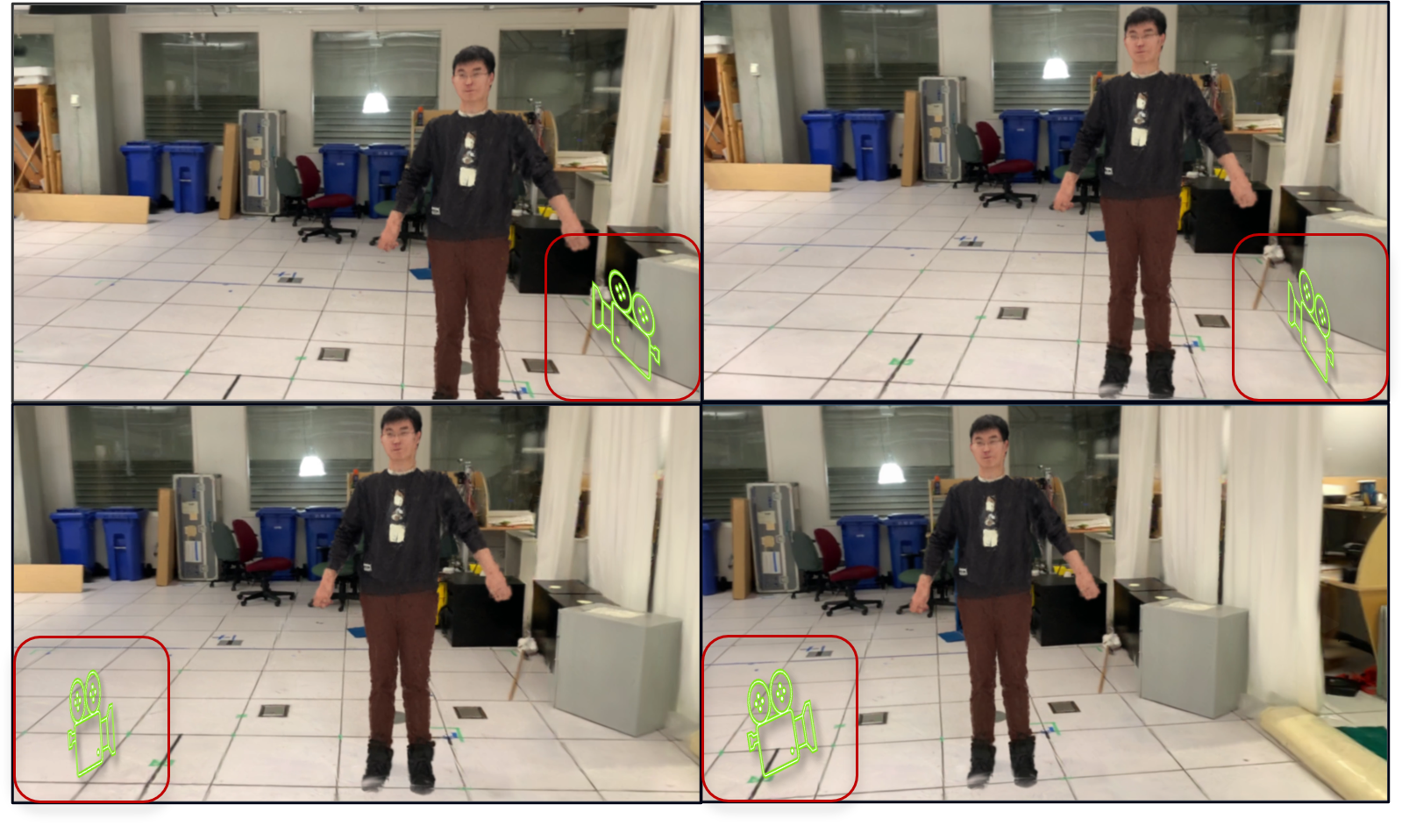}
  \caption{Motion editing in real-time in different views.}
  \label{fig:motion_edit__view}
\end{figure}

The Figure~\ref{fig:motion_edit} and ~\ref{fig:motion_edit__view} illustrate real-time editing of motion. We can directly edit individual joints of the human body to simulate interactive motion editing. For instance, in Figure~\ref{fig:motion_edit}, we modify the joints, demonstrating the precise editing capabilities of our method. Additionally, our system enables real-time feedback, allowing users to adjust the human pose while seamlessly integrating it with the background. In Figure~\ref{fig:motion_edit__view}, we can also change the view during the motion editing. This feature is handy for animation creation and motion correction tasks.

\subsection{Novel Motion Synthesis}
\begin{figure}[h]
  \centering
  \includegraphics[width=\linewidth]{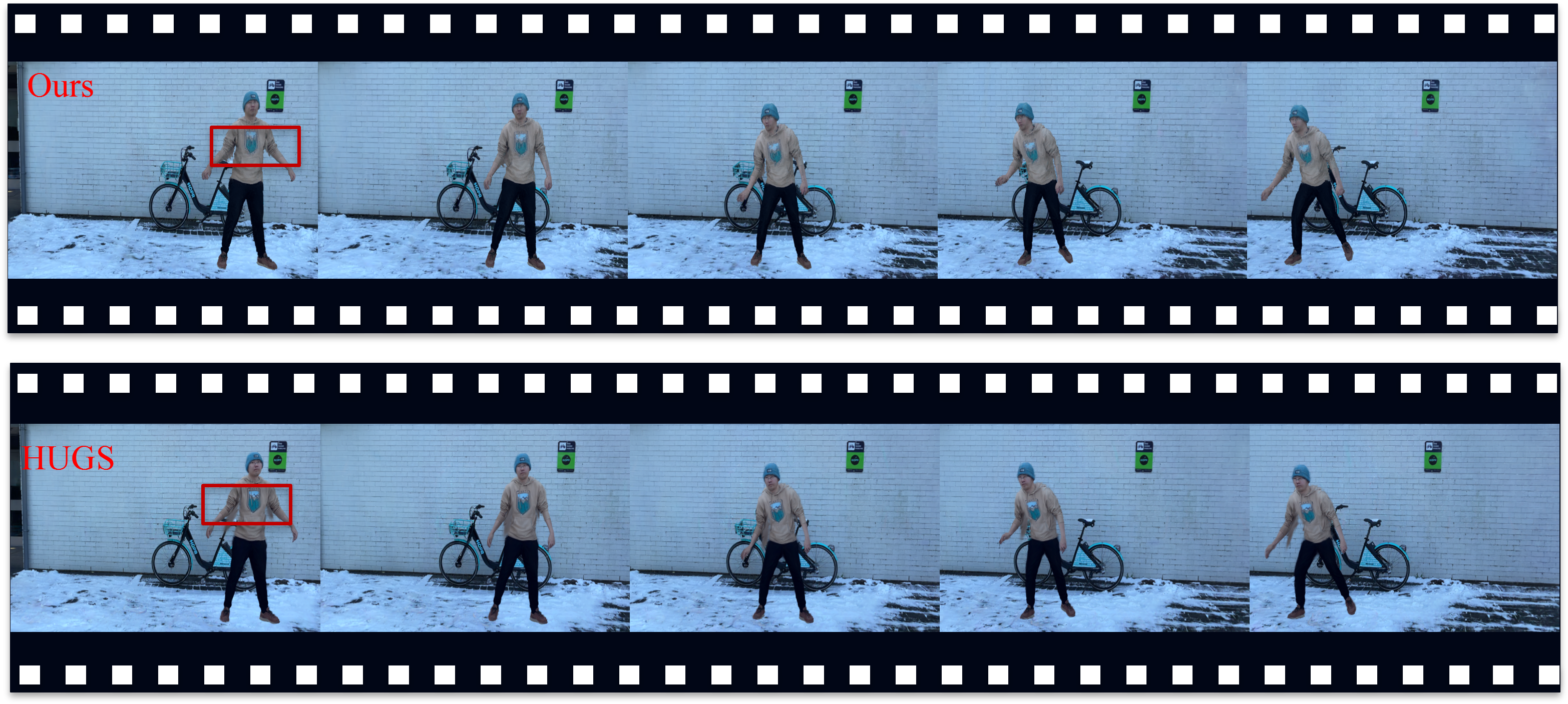}
  \caption{Synthesizing novel motions using our method (above) and HUGS (below).}
  \label{fig:new_motion}
\end{figure}

\begin{figure}[h]
  \centering
  \includegraphics[width=\linewidth]{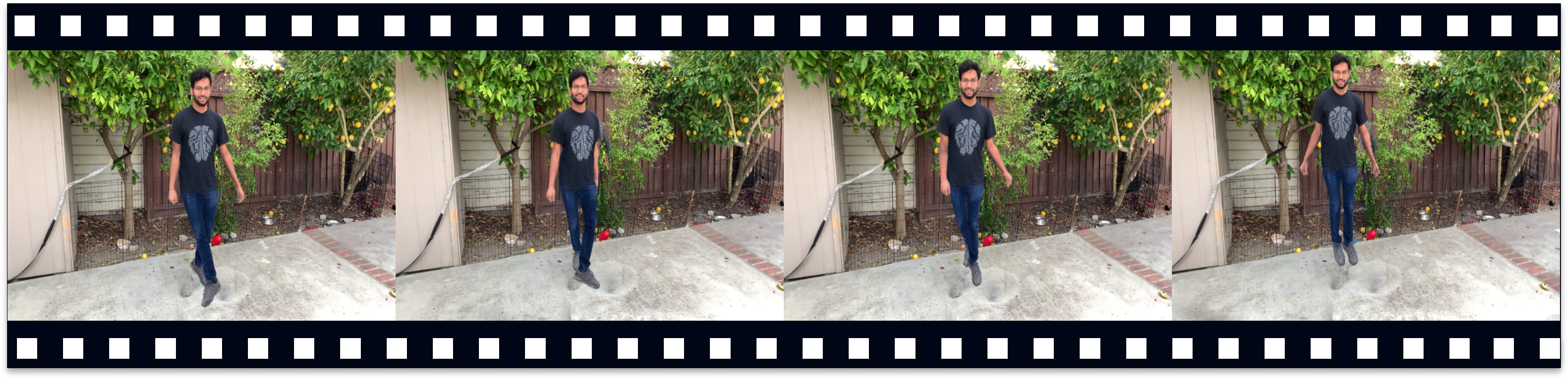}
  \caption{Synthesizing novel motions using our method.}
  \label{fig:new_motion_2}
\end{figure}

Another application is generating new motion. Using the edited motion, we can apply them to the scene and simulate dynamic interactions between the human body and its surroundings. As shown in Figures~\ref{fig:new_motion} and ~\ref{fig:new_motion_2}, different motion sequences (e.g., walking and dancing) are applied on a reconstructed background.

\subsection{Animal Scene Reconstruction}

By leveraging advanced platforms like the SMPL human model, we demonstrate the versatility of our framework in handling scene-based motion applications. This is not limited to human motion but extends seamlessly to animal motion reconstruction through the SMAL model~\cite{zuffi20173d}.

\begin{figure}[]
  \centering
  \includegraphics[width=\linewidth]{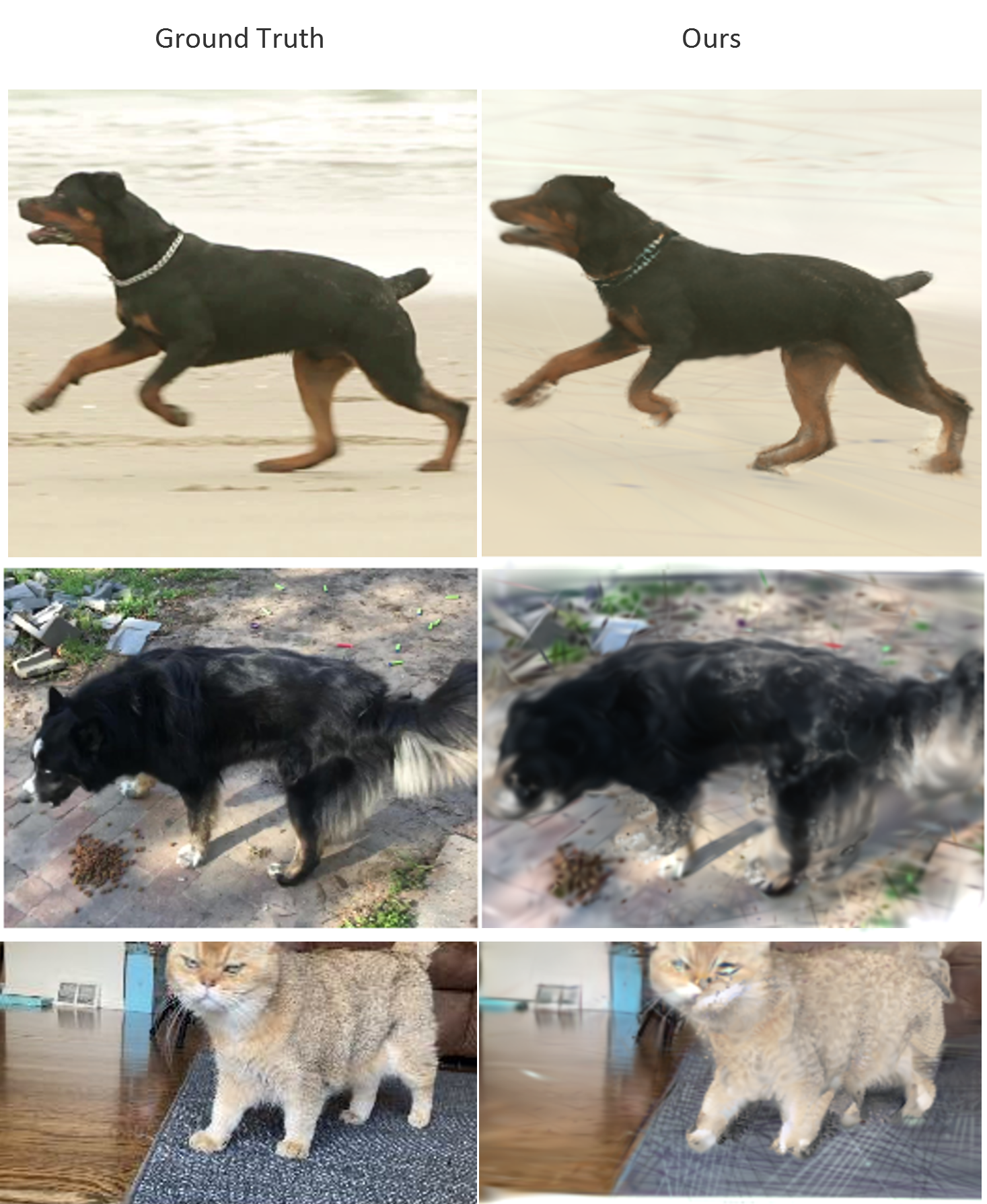}
  \caption{Animal reconstruction for quadruped in different scenarios.}
  \label{fig:animal_app}
\end{figure}

\begin{figure}[h]
  \centering
  \includegraphics[width=\linewidth]{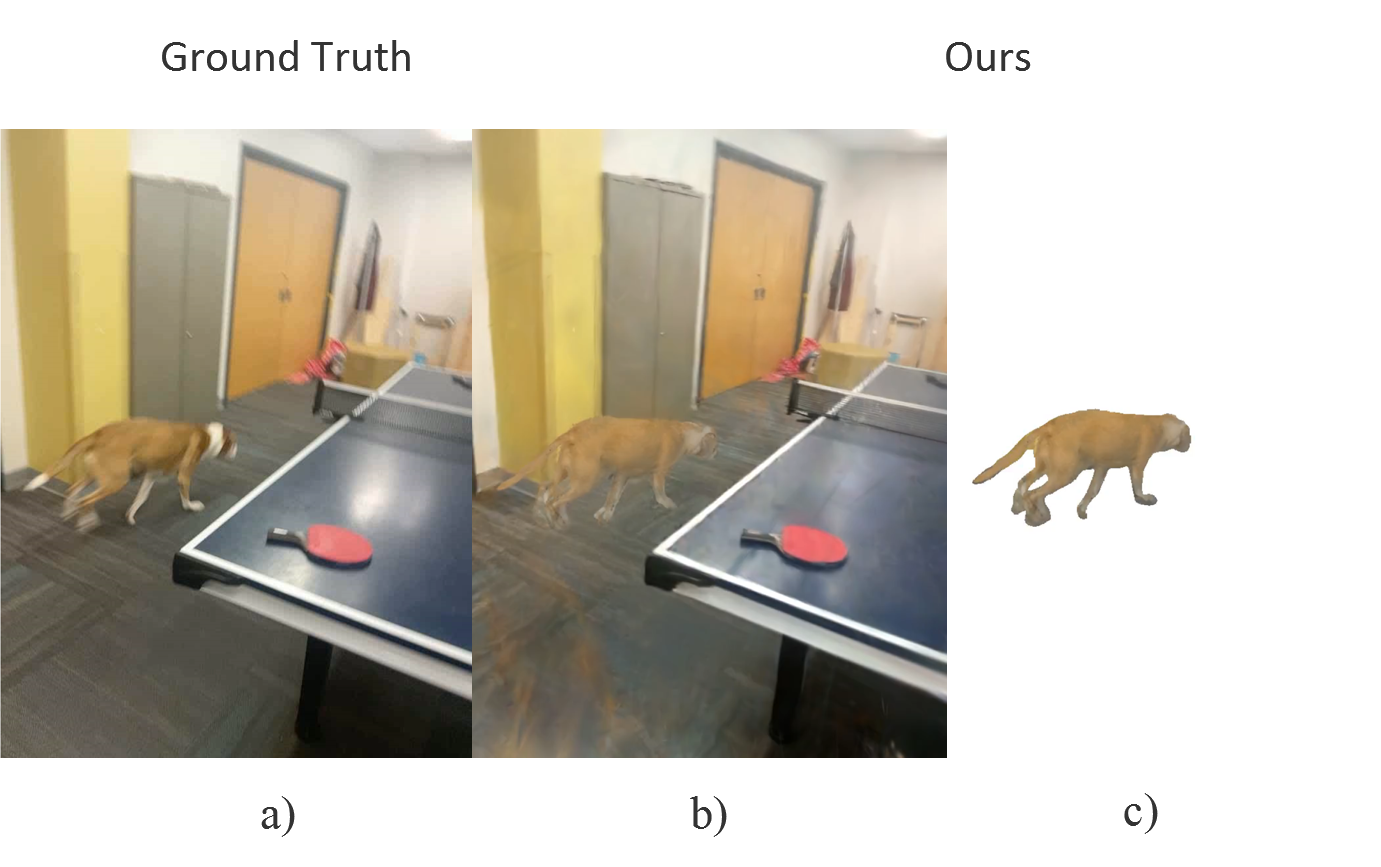}
  \caption{The reconstruction of one dog sample.}
  \label{fig:animal_app_2}
\end{figure}

Figures~\ref{fig:animal_app} and~\ref{fig:animal_app_2} showcase how the SMAL model enables the reconstruction of complex animal motion, providing an editing effect comparable to human motion applications. This capability highlights the adaptability and robustness of our method in diverse scenarios.
In Figure~\ref{fig:animal_app_2}, a) shows the ground truth, b) demonstrates the complete reconstruction effect, where both the shape and motion of the animal are captured with reasonable fidelity, and c) shows the effect of animal segmentation. The datasets used for these experiments include Smalift~\cite{biggs2019creatures}, Cop3d~\cite{sinha2023common}, and Rac~\cite{yang2023reconstructing}.

Our approach combines the SMPL and SMAL models, enabling smooth transitions between human and animal motion modeling. This innovation paves the way for creating realistic and coherent animations and exploring the underlying mechanics of animal motion in scientific studies.

\section{Conclusion}
We propose a novel approach for 3D human dynamic scene reconstruction based on a skinning model, effectively disentangling humans and backgrounds. Our method leverages convolutional networks to predict human Gaussian point features and real-time rendering fields, enhancing scene realism. By introducing a method for point cloud growing based on a position texture, we achieve richer body representation and detail capture while using fewer points than HUGS, enabling rendering speeds exceeding 100 FPS.

Furthermore, our texture mapping-based feature predictor ensures appearance consistency during background co-optimization, resolving occlusion issues in human-scene contact areas. Our method surpasses HUGS and Neuman in reconstruction quality and extends to quadruped animal reconstruction by replacing the SMPL model with SMAL, offering potential applications in behavioral studies and 3D asset creation for animal scenes.

\textbf{Limitations and Future Work.} While our method effectively reconstructs dynamic human scenes from monocular videos, challenges persist in pose estimation, clothing representation, and multi-person scenarios. Pose accuracy depends on SMPL parameters from ROMP, and while Neuman's re-optimization mitigates errors, integrating Gaussian-based refinement could further enhance precision. The current framework cannot model loose-fitting clothing, but incorporating clothing features and LBS-driven parameters, as in ASH~\cite{pang2024ash}, presents a promising direction. Additionally, our approach struggles with multi-person dynamic scenes, necessitating future work on disentangling interactions and integrating scene-specific predictors for individual humans.

\section*{Acknowledgments}
We extend our gratitude to the developers of the outstanding open-source projects COLMAP, ROMP, Gaussian Splatting, Gaussian Avatar, Neuman, and HUGS, whose contributions have significantly supported the development of our work. We also sincerely thank everyone who provided valuable feedback and assistance throughout this project.

\bibliographystyle{unsrt}  
\bibliography{main}

\end{document}